\documentclass[aps,prb,a4paper,twocolumn]{revtex4}

\usepackage{amsmath}
\usepackage{amssymb}
\usepackage{graphicx}
\usepackage{epsfig}
\usepackage{bbm}
\usepackage{color}
\usepackage[cm]{fullpage}

\usepackage[ps2pdf,
bookmarks=true,
bookmarksnumbered=true,
hypertexnames=false,
]{hyperref}

\newcommand{\bra}[1]{\left< #1 \right|}
\newcommand{\ket}[1]{\left| #1 \right>}

\newcommand{\braket}[2]{\left<\left.\! #1 \right|\! #2 \right>}

\newcommand{\mytitle}{Introducing many-body physics using atomic spectroscopy} 

\newcommand{\rmpdfinfo}{\special{ps:: userdict /pdfmark /cleartomark load put}}

\definecolor{MyDarkGreen}{rgb}{0,0.6,0}
\definecolor{MyDarkBlue}{rgb}{0,0,0.8}
\definecolor{MyDarkRed}{rgb}{0.6,0,0.3}

\hypersetup{breaklinks=true,
colorlinks =true,
plainpages =true,
linktocpage=true,
linkcolor=MyDarkBlue,
citecolor=MyDarkGreen,
urlcolor =MyDarkRed,
pdfborder={0 0 0},
pdfauthor={Dietrich Krebs, Stefan Pabst, and Robin Santra},
pdftitle ={\mytitle},
pdfsubject  = {Research Article},
pdfkeywords = {TDCIS},
pdfcreator  = {LaTeX with hyperref package},
pdfproducer = {dvipdf}
}

\begin{document} 

\title{\mytitle}

\author{Dietrich Krebs}
\affiliation{Center for Free-Electron Laser Science, DESY, Notkestrasse 85, 22607 Hamburg, Germany}
\affiliation{Department of Physics,University of Hamburg, Jungiusstrasse 9, 20355 Hamburg, Germany}

\author{Stefan Pabst}
\affiliation{Center for Free-Electron Laser Science, DESY, Notkestrasse 85, 22607 Hamburg, Germany}
\affiliation{Department of Physics,University of Hamburg, Jungiusstrasse 9, 20355 Hamburg, Germany}

\author{Robin Santra}
\email{robin.santra@cfel.de}
\affiliation{Center for Free-Electron Laser Science, DESY, Notkestrasse 85, 22607 Hamburg, Germany}
\affiliation{Department of Physics,University of Hamburg, Jungiusstrasse 9, 20355 Hamburg, Germany}

\date{\today}

\begin{abstract}
Atoms constitute relatively simple many-body systems, making them suitable objects for developing an understanding of basic aspects of many-body physics.
Photoabsorption spectroscopy is a prominent method to study the electronic structure of atoms and the inherent many-body interactions.
In this article the impact of many-body effects on well-known spectroscopic features such as Rydberg series, Fano resonances, Cooper minima, and giant resonances is studied, and related many-body phenomena in other fields are outlined.
To calculate photoabsorption cross sections the time-dependent configuration interaction singles (TDCIS) model is employed. The conceptual clearness of TDCIS in combination with the compactness of atomic systems allows for a pedagogical introduction to many-body phenomena. 
\end{abstract}
  
  
\maketitle

\section{Introduction}
\label{sec:intro}
%
%
Atomic systems have been studied for more than a century, inspiring developments from early quantum mechanics\cite{Bo-1913-Atommodell_Bohr} to modern attosecond physics.\cite{KrIv-2009-RevModPhys.81.163_Atto_REVIEW}
They continue to be at the focus of current research,\cite{KiGo-2004-nature_transient_recorder_streaking} especially as they constitute comparatively simple many-body systems.\cite{Am-1990-BOOK_atomic_photoeffect,LiMo-1986-BOOK_Atomic_many_body_theory} Developing an understanding of these systems helps to further comprehend many-body effects in more complex environments such as ultracold gases\cite{BlDa-2008-RevModPhys.80.885_Ultracold_Many_Body_REVIEW} or condensed matter phases.\cite{Fu-2002-BOOK_Condensed_matter_many_body}

The purpose of this article is to highlight atomic physics as a teaching tool for the introduction of many-body physics. 
In our experience, many-body physics is discussed rather late in physics curricula. 
To facilitate an earlier introduction, we advocate using atomic physics to present to students some of the most basic many-body phenomena in an elementary way. 
Such an introduction can already be taught in undergraduate courses, serving later on as a valuable foundation for more advanced courses. 

In order to present elementary many-body physics in a way suitable for both experimental and theoretical courses, we have chosen prominent phenomena from atomic spectroscopy (i.e., Rydberg series,
Fano profiles,
Cooper minima,
and giant resonances) as examples for illustrating the influence of many-body effects (see Section~\ref{sec:spec_gen}).
We employ the photoabsorption cross section as a characteristic quantity, because it is experimentally accessible through the widely used technique of photoabsorption spectroscopy\cite{WaOk-2012-Photoabsorption_spectr_on_GaAs_excitons,LoJo-2012-photoabsorption_geminal_ethylene_difluoride} 
and has been thoroughly studied especially for noble gases.\cite{SaSt-2002-JoElSp2002265}

To give a theoretical background and enable a systematic analysis, we provide an intuitive theoretical model to teach the light-matter interaction and many-body effects in noble gas atoms.
This model is essentially a time-dependent formulation of the widely used configuration-interaction-singles (CIS) approach.\cite{SzaboOstlund_ModernQuantumChemistry,FoPo-1992_JPhysChem_CIS} 
We describe the state of the atomic system in terms of its Hartree-Fock ground state and particle-hole excitations thereof, maintaining at the same time the full, nonrelativistic Hamiltonian in order to capture electronic correlation dynamics (see Section~\ref{sec:theory}).
With this approach we aim to preserve the explanatory simplicity of the successful independent electron picture (see, e.g., Ref.~\onlinecite{CooperNATO1975}) and offer insight into electronic many-body effects at the same time.
The implementation of the time-dependent configuration interaction singles (TDCIS) method is given in detail in Refs.~\onlinecite{RoSa-2006-PhysRevA.74.043420} and~\onlinecite{GrSa-2010-PhysRevA.82.023406}.
The method has been used to investigate optical strong-field processes,\cite{Pa-2013-EPJ_ST_phd_as_article} having found successful application to high-harmonic generation,\cite{PaSa-2012-PhysRevA.85.023411} strong-field ionization,\cite{PaSy-2012-PhysRevA.86.063411_transient_absorption} and intense ultrafast x-ray physics.\cite{SySa-2012-PhysRevA.85.023414}
In this article we demonstrate that it also captures essential parts of experimentally observed photoabsorption spectra.\cite{SaSt-2002-JoElSp2002265} 
Although we are not presenting novel physics with these results, we believe they are very suitable for the classroom, because they combine interesting many-body effects with an intuitive explanatory model.

While its performance and its comparably simple ansatz recommend TDCIS as a teaching model, 
note that the inherent picture of particle-hole excitations finds more general applications in a variety of many-body systems beyond atoms. For example, in nuclear physics,\cite{MaSa-1969-Nucl.Phys_GDR_1p1h} as well as in solid-state physics,\cite{Al-1996-Collective_Excitations} the Tamm-Dancoff approximation (which is comparable to CIS) can be used to describe collective excitation phenomena.

As outlined, the theoretical framework of TDCIS helps to provide an understanding of many-body phenomena in atomic systems. However, depending on the course level and topic, one could skip parts of the theoretical section and concentrate more on the spectroscopic features. In any case we recommend alluding to the concept of particle-hole excitations, which appears in almost all areas of many-body physics. Suggestions for illustrative additional topics---both theoretical and experimental---are indicated and connected to the aforementioned phenomena.

\section{Theoretical Background}
\label{sec:theory}
\label{sec:theo_over}

In order to investigate the many-body phenomena that are imprinted in the atomic photoabsorption cross section, it is essential to have an understanding of both the atomic system and its interaction with an electromagnetic field. We therefore proceed by presenting the Hamiltonian $\hat H$ of the system in Section~\ref{sec:theo_hamiltonian}, before we discuss the atomic state in terms of its wave function $ \ket{\Psi(t)}$ in Section~\ref{sec:theo_TDCISwavefct}. The actual photoabsorption process is then described in Section~\ref{sec:theo_TDSE} by solving the time-dependent Schr\"odinger equation (TDSE).

\subsection{Hamiltonian}
\label{sec:theo_hamiltonian}
To describe an atomic system and its photoabsorption behavior, we employ the exact nonrelativistic Hamiltonian, incorporating the interaction of the atom with a classical electromagnetic field according to the principle of minimal coupling.\cite{Sa-2009-J.Phys.B_phd_tutorial_XRAY,CrTh-1998-BOOK_Molecular_Quantum_Electrodynamics_minimal_coupling} With both the Coulomb gauge and the dipole approximation\cite{Starace-HandbuchPhysRadiation1982} imposed upon the electromagnetic field, the Hamiltonian reads
\begin{align}
  \label{eq:Hamiltonian2}
    \hat H
  &=
  \sum_{n} \left[ \frac{{\vec{p}}_n^{\,\,2}}{2} - \frac{Z}{\left|\vec{r}_n\right|} \right]
    + \frac{1}{2}\! \sum_{n \ne m} \frac{1}{\left|\vec{r}_n - \vec{r}_m\right|}
    + \sum_{n} {A}(t){(\hat{p}_{z})_n}
,
\end{align}
where $\vec{p}_n$ and $\vec{r}_n$ are the momentum and position operators, respectively, for the $n$th electron, with the $\hat{\phantom{o}}$ denoting an operator omitted from vectors to avoid notational clutter.
Here we assume the vector potential $\vec{A}(t)$ to be linearly polarized along the $z$ axis in order to simplify further calculations. Note that we employ atomic units throughout this article to keep expressions conveniently concise (i.e., $m_e = \left|e\right| = \hbar = 1/(4\pi\varepsilon_0) = 1$, where $m_e$ is the electron mass, $e$ its charge, $\hbar$ the reduced Planck constant, and $\varepsilon_0$ the permittivity of free space).

In Eq.~\eqref{eq:Hamiltonian2} the summation indices $n$ and $m$ run over the number of electrons in the system, ranging from $1$ to $N$. 
The first sum comprises the total electron kinetic energy and the attractive central potential of the nucleus (charge $Z$) in which the $N$ electrons move. Note that we assume this nucleus to be infinitely heavy. The second summation incorporates the repulsive Coulomb interaction among the electrons, which differs decidedly from the aforementionned single-body contributions. 
It is this two-body Coulomb term that gives rise to many-body phenomena and poses the decisive challenge in many-body-calculations. Due to its inherent complexity, the Coulomb interaction is often just approximated by an effective single-body potential.\cite{SzaboOstlund_ModernQuantumChemistry}
The third term in Eq.~\eqref{eq:Hamiltonian2} is again composed of single-body operators and describes the coupling of each electron to the electromagnetic field.

In a first approach it proves advantageous to approximate Eq.~\eqref{eq:Hamiltonian2} as follows. In the absence of an electromagnetic field ($\vec{A}(t)=0$), the atomic system can be described by an approximate mean-field Hamiltonian $\hat{H}_0$:
\begin{align}
  \label{eq:HamiltonianMF}
    \hat{H}_0
  &=
  \sum_{n=1}^N \left[ \frac{\vec{p}_n^{\,\,2}}{2} - \frac{Z}{\left|\vec{r}_n\right|} + \hat{V}^{MF}_n \right]
   .
\end{align}
This implies that the interaction among the electrons is reduced to a state in which each electron ``senses'' only a mean potential $\hat{V}^{MF}_n$ produced by the other $N-1$ electrons.
The key point is that Eq.~\eqref{eq:HamiltonianMF} contains only single-body operators. 
In order to recover the full many-body Hamiltonian (Eq.~\eqref{eq:Hamiltonian2}) from Eq.~\eqref{eq:HamiltonianMF} we have to add to $\hat{H}_0$ both the interaction with the electromagnetic field and a correction potential $\hat{V}^{ee}$ that contains the difference between the full Coulomb interaction and the mean-field potential.
In doing so, we obtain
\begin{align}
  \label{eq:Hamiltonianfinal}
   \hat{H}
  &=
    \hat{H}_0 + \hat{V}^{ee} + A(t)\hat{P} - E_0^{MF}
.
\end{align}
Here we have used the symbol $\hat{P}=\sum_{n=1}^N {(\hat{p}_{z})_n}$ and applied a global energy shift by the mean-field ground-state energy $E_0^{MF}$, which serves purely cosmetic purposes. Partitioning the Hamiltonian in the outlined fashion simplifies the later solution of the TDSE and allows for a particularly straightforward control of many-body correlations by manipulation of $\hat{V}^{ee}$.

\subsection{TDCIS wave function}
\label{sec:theo_TDCISwavefct}

In constructing a wave-function ansatz for the description of the atomic state, we proceed in two steps.
In the first step we construct the ground state of the system based on a mean-field theory and in the second step we add excited configurations.

The field-free ground state of a closed-shell system, such as a noble gas atom, can be well approximated by a Hartree-Fock (HF) calculation.\cite{SzaboOstlund_ModernQuantumChemistry} This assumes that each electron occupies a single particle spin orbital $\ket{\varphi_p}$, sensing only the mean field of the other $N-1$ electrons.
If $\hat{H}_0$ (see Eq.~\eqref{eq:HamiltonianMF}) is chosen to be the Fock operator,\cite{SzaboOstlund_ModernQuantumChemistry} these orbitals $\ket{\varphi_p}$ are generated in a self-consistent fashion alongside the HF potential $\hat{V}^{MF}_n$ by solving the eigenvalue equation
\begin{align}
  \label{eq:Fock}
    \hat{H}_0 \ket{\varphi_p}
  &=
  \varepsilon_p \ket{\varphi_p}
   .
\end{align}
Associated with each orbital $\ket{\varphi_p}$ is the orbital energy $\varepsilon_p$.
For further details on the HF procedure, see Refs.~\onlinecite{SzaboOstlund_ModernQuantumChemistry} and~\onlinecite{EcAl-2007-doi:10.1080/00268970701757875_HFReview}.
Taking the antisymmetrized product (Slater determinant) of the $N$ energetically lowest spin orbitals yields the HF ground state $\ket{\Phi_0}$:
\begin{align}
  \label{eq:ground state}
    \ket{\Phi_0}
  &=
  \det{\left(\ket{\varphi_1},...,\ket{\varphi_N}\right)}
.
\end{align}
This state will serve as a reference state to build a many-body theory that reaches beyond the mean-field picture. Various other many-body approaches, referred to as post-HF methods,\cite{SzaboOstlund_ModernQuantumChemistry,ShBa-2009-BOOK-Many_Body_Methods_in_Chemistry_and_Physics} share this starting point. To account for electronic excitations we add to our wave function singly excited configurations $\ket{\Phi^a_{i}}$, in which an electron from the \mbox{$i$th} orbital $\ket{\varphi_i}$ (occupied in $\ket{\Phi_0}$) is promoted into the \mbox{$a$th} orbital $\ket{\varphi_a}$ (unoccupied in $\ket{\Phi_0}$):
\begin{align}
  \label{eq:excitedstate}
    \ket{\Phi^a_{i}}
  &=
  \det{\left(\ket{\varphi_1},...,\ket{\varphi_{i-1}},\ket{\varphi_a},\ket{\varphi_{i+1}},...,\ket{\varphi_N}\right)}
.
\end{align}
The excited atomic system may be pictured to exhibit a positively charged ``hole'' (the vacant orbital $\ket{\varphi_i}$) which
characterizes the state of the \mbox{$N-1$} electrons which remain unexcited.
The excited (possibly ionized) atomic state arising from a photoabsorption process is often characterized in terms of the hole's quantum numbers, which define the excitation (or ionization) \emph{channel} (see, e.g., Ref.~\onlinecite{SuPe-2012-JPhysB_REVIEW_AutoIonRes_raregases}).

Describing a system's wave function using the ground state and single excitations is known as \textit{configuration interaction singles} (CIS). One can include further excitation classes beyond singles, e.g., double excitations $\ket{\Phi^{a_1,a_2}_{i_1,i_2}}$. Considering all excitation classes up to $N$---yielding $\ket{\Phi^{a_1,a_2,...,a_N}_{i_1,i_2,...,i_N}}$---is referred to as full CI (FCI).\cite{SzaboOstlund_ModernQuantumChemistry}

We will, however, limit our wave function to the CIS ansatz, which preserves the intuitive picture that the hole index $i$ refers to the ionic eigenstate. (This is not valid for higher-order CI methods.\cite{BrCe-2003-hole_migration_molecules_JCP})
Using time-dependent expansion coefficients $\alpha_{0}(t)$ and $\alpha^a_{i}(t)$, our wave-function ansatz reads:
\begin{align}
  \label{eq:TDCISwavefct}
  \ket{\Psi(t)}
  &=
  \alpha_0(t) \ket{\Phi_0}
  +
  \sum_{a,i} \alpha^a_{i}(t) \ket{\Phi^a_{i}}
  .
\end{align}
Though only single excitations are accessible, their coherent superpositions can capture a decisive share of the electronic many-body correlations and collective excitations. This quality is reflected by the good agreement between theoretical and experimental results (see Section~\ref{sec:spec_gen}). The influence of many-body effects beyond our wave-function ansatz can also be probed within the presented theoretical framework, as we outline in Appendix~\ref{app:forms}.

\subsection{The photoabsorption process}
\label{sec:theo_TDSE}
In order to investigate the process of photoabsorption we study the temporal evolution of the wave function $\ket{\Psi(t)}$. To do so, we have to solve the time-dependent Schr\"odinger equation (TDSE):
\begin{align}
  \label{eq:TDSE}
  i\frac{\partial}{\partial t} \ket{\Psi(t)}
  &=
  \hat H \ket{\Psi(t)}
  .
\end{align}
We obtain the explicit expression of the TDSE by inserting both our wave function ansatz (Eq.~\eqref{eq:TDCISwavefct}) and the Hamiltonian presented in Eq.~\eqref{eq:Hamiltonianfinal}:
\begin{align}
  \label{eq:TDSE2}
  i\dot\alpha_0(t) \ket{\Phi_0}
  \!+\!
  i\sum_{a,i} \dot\alpha^a_{i}(t) \ket{\Phi^a_{i}}
  \!=\!\!  
  \left[\hat{H}_0 \!+\! \hat{V}^{ee} \!-\! A(t)\hat{P} \!-\! E_0^{MF}\right]&
  \nonumber\\
  \times
  \Big(\alpha_0(t) \ket{\Phi_0}
  +
  \sum_{a,i} \alpha^a_{i}(t) \ket{\Phi^a_{i}}
  \Big)
  &.
\end{align}
In a further step we project Eq.~\eqref{eq:TDSE2} on the basis states $\bra{\Phi_0}$ and $\bra{\Phi^a_{i}}$ (for details see Ref.~\onlinecite{RoSa-2006-PhysRevA.74.043420}) and obtain a set of equations of motion for the expansion coefficients $\alpha_{0}(t)$ and $\alpha^a_{i}(t)$:
\begin{subequations}
\label{eq:eom}
\begin{align}
\label{eq:eom_alpha0}
  i\,\dot\alpha_0(t)
  &=
  A(t) \sum_{i,a} 
  \bra{\Phi_0}\hat P\ket{\Phi^a_{i}}
  \alpha^a_{i}(t) 
\\
\nonumber
  i\,\dot\alpha^a_{i}(t)
  &=
  (\varepsilon_a-\varepsilon_i) \, \alpha^a_{i}(t)
  +
  \sum_{b,j}
  \bra{\Phi^a_{i}}\hat{V}^{ee}\ket{\Phi^b_{j}}
  \alpha^b_{j}(t)
\\\label{eq:eom_alpha_ai}
  & + A(t)\!\Big[
  \!\bra{\Phi^a_{i}}\hat P\ket{\Phi_0}
  \alpha_0(t)
  +
  \!\sum_{b,j}
  \bra{\Phi^a_{i}}\hat P\ket{\Phi^b_{j}}
  \alpha^b_{j}(t) 
  \Big]
  .
\end{align}
\end{subequations}
Here the dipole matrix elements $\bra{\Phi_0}\hat P\ket{\Phi^a_{i}}$ describe the initial photoexcitation of an electron from an occupied into an unoccupied spin orbital (from $\ket{\varphi_i}$ into $\ket{\varphi_a}$). The dipole transitions $\bra{\Phi^a_{i}}\hat P\ket{\Phi^b_{j}}$ correspond to the subsequent absorption of photons. 
These are, however, irrelevant for the one-photon processes that are of primary interest in this paper.

While $\hat P$ is only a single-body operator, the residual electron-electron interaction, $\hat{V}^{ee}$, couples two particles---the excited electron and the hole.
This so called interchannel coupling can appear in two distinct ways.  The first is referred to as direct interaction, which is the relocation of the excited electron from an orbital $\ket{\varphi_a}$ into an orbital $\ket{\varphi_b}$ while \emph{simultaneously} 
the hole moves from $\ket{\varphi_i}$ into $\ket{\varphi_j}$.
The second possibility arises from the antisymmetric nature of the wave function and is known as exchange interaction,\cite{SzaboOstlund_ModernQuantumChemistry} where an excited electron ($\ket{\varphi_a}$) recombines with the hole $\ket{\varphi_i}$ and in exchange another electron is excited from $\ket{\varphi_j}$ into $\ket{\varphi_b}$, thereby creating a new electron-hole pair.
These two processes are analogous to the two channels that contribute to Bhabha scattering of a positron (hole) and an electron in particle physics.\cite{PeSc-1995-Introduction_QFT_Bhabha}

To investigate the impact of these electronic interactions, we will compare the results of \emph{full} TDCIS calculations, which allow for all electronic couplings $\bra{\Phi^a_{i}}\hat{V}^{ee}\ket{\Phi^b_{j}}$,
with the more restricted \emph{intrachannel} calculations, for which we artificially enforce $\bra{\Phi^a_{i}}\hat{V}^{ee}\ket{\Phi^b_{j}} = 0$ if $i \ne j$.
In the latter case an electron, once it is excited, cannot change the ionic state.  It senses the effective potential of the residual ion, but does not partake in many-body correlations.

Using states that are based on HF orbitals leads to the interesting and useful side effect that $\hat{V}^{ee}$ does not couple the ground state $\ket{\Phi_0}$ to singly excited configurations $\ket{\Phi^a_{i}}$. This is known as Brillouin's theorem.\cite{SzaboOstlund_ModernQuantumChemistry} It allows for a straightforward study of photoabsorption, as the atomic ground state, which is initially occupied (i.e., $\alpha_{0}(t)=1$ for $t\to - \infty$), can only be depopulated by dipole transitions in an electromagnetic field.
In the absence of the field, the ground-state amplitude $\alpha_0(t)$ remains unchanged, as Eq.~\eqref{eq:eom_alpha0} reduces to $ i\dot\alpha_0(t)=0$. This implies that the change of $\left| \alpha_0 \right|^2$ during a light pulse can directly be related to photoabsorption.
For a light pulse of perturbatively weak intensity, the one-photon photoabsorption cross section\cite{endnote1} ($\sigma$) can be calculated directly as
\begin{align}
  \label{eq:alphacross}
  \sigma
  &=
  \frac{1-\left| \alpha_0 \right|^2}{N_\gamma}
  ,
\end{align}
where $N_\gamma$ is the number of photons per unit area in the incident light pulse.\cite{endnote2} 
Note that Eq.~\eqref{eq:alphacross} yields the cross section $\sigma$ corresponding to only a single pulse.
In order to record a full photoabsorption spectrum $\sigma(\omega)$ we have to perform multiple simulations, employing pulses at different mean photon energies.
The resolution of this method is then set by the energy spacing between the pulses and their spectral width. 
We present an alternative to this numerically expensive approach in Appendix~\ref{app:cross}.

\section{Spectroscopic features in TDCIS}
\label{sec:spec_gen}
\label{sec:spec_overview}
\begin{figure*}[ht!]
\begin{center}
  \rmpdfinfo
  \includegraphics[width=.8\textwidth]{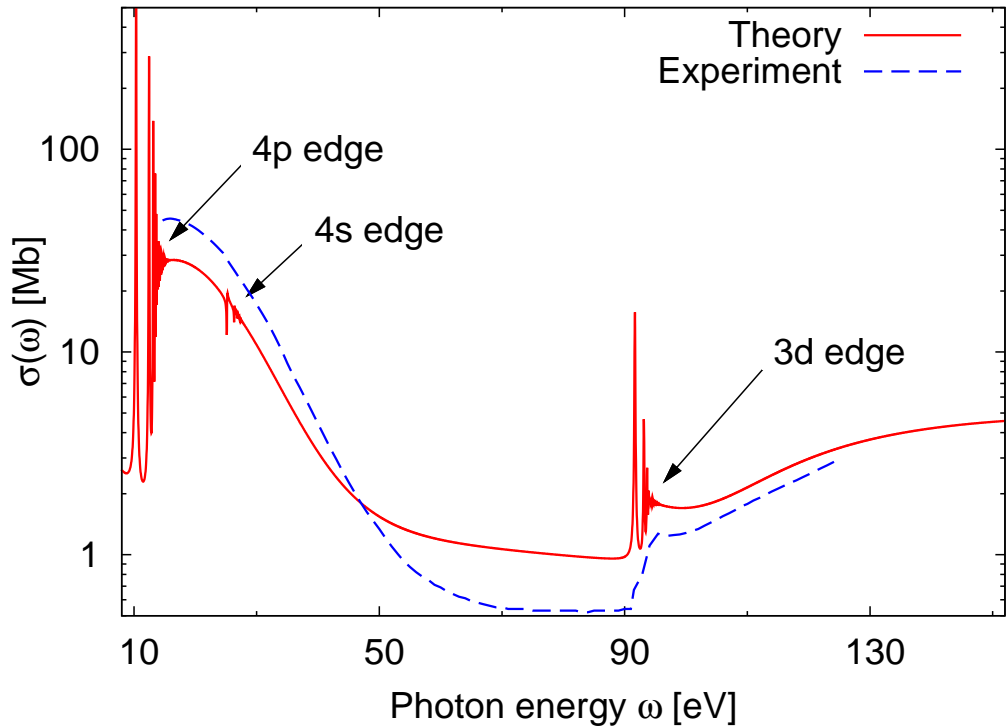}
  \caption{(Color online.) The calculated photoabsorption cross section $\sigma(\omega)$ of krypton (solid line) is plotted as a function of the incident photon energy $\omega$. It is compared to the experimental results (dashed line) from Ref.~\onlinecite{SaSt-2002-JoElSp2002265}. Ionization edges of outer subshells are marked by arrows. 
  }
  \label{fig:krover}
\end{center}
\end{figure*}

A powerful tool to investigate the electronic structure of atomic systems is the photoabsorption cross section.
It encodes a wide range of phenomena which find their respective analogues in various other areas, like solid-state or ultracold physics. 
The theoretical cross sections presented in this article were calculated with the \textsc{xcid} program package,\cite{endnote3} which implements the TDCIS model.\cite{endnote4}
A typical cross section over a large energy range is shown in Fig.~\ref{fig:krover} for atomic krypton.
On first sight one recognizes distinctive structures preceding the ionization threshold of each subshell. (The threshold locations are\cite{XDB} $14.1$~eV, $27.5$~eV, and $93.8$~eV, for $4p$, $4s$, and $3d$, respectively.)
These line spectra are referred to as Rydberg series (see Section \ref{sec:spec_rydberg}).
For photon energies above a threshold, the cross section typically declines in a monotonic fashion, 
until the next subshell becomes energetically accessible.
Embedded in the $4p$ continuum lies the $4s$ Rydberg series of krypton which will---on closer examination---show a modified line profile, reflecting electronic correlation (see Section~\ref{sec:spec_fano}).
Following the $3d$ edge the cross section rises to form an extended resonance in the continuum. This prominent effect of electron-electron interaction will further be studied in Section~\ref{sec:spec_gdre} in connection with the $4d$ giant resonance of xenon.
In Section~\ref{sec:spec_coop} we will examine the cross section of argon and discuss what is known as a Cooper minimum.

\subsection{Rydberg series}
\label{sec:spec_rydberg}

We now turn our attention more closely to the line spectra preceding ionization edges.
They owe their name ``Rydberg series'' to J. Rydberg, who in 1889 presented\cite{Ry-1889-Rydberg_formula} the well-known parametrization of the line spectrum of hydrogen:
\begin{align}
  \label{eq:rydbergH}
  E_{n}
  &=
  \frac{1}{2} - \frac{1}{2\,n^2}
  ,
\end{align}
where $E_n$ is the energy required to excite the electron in the Coulomb potential of the proton ($V(r)=-1/r$) from its ground state ($n = 1$) into an excited, yet bound, state with principal quantum number~$n$.
Every bound-to-bound transition gives rise to a distinct peak in the photoabsorption spectrum, jointly forming a series of lines, with their positions converging up to the ionization threshold $I_{p}=1/2$.

For larger atomic species such line spectra will likewise emerge if an electron is excited in the nuclear potential ($-Z/r_n$), but the positions of the lines will be strongly influenced by the Coulomb interaction of the excited electron with the remaining $N-1$ electrons. 
In the TDCIS model we picture the properties of such an excited atom in terms of the excited electron and its corresponding hole.
Therefore we can interpret Rydberg states essentially as the bound states of an electron-hole pair, an approach that is analogous to the treatment of excitons in solid state physics.\cite{Li-1970-PhysEduc_Excitons}

\begin{figure*}[ht!]
\begin{center}
  \rmpdfinfo
  \includegraphics[width=.8\textwidth]{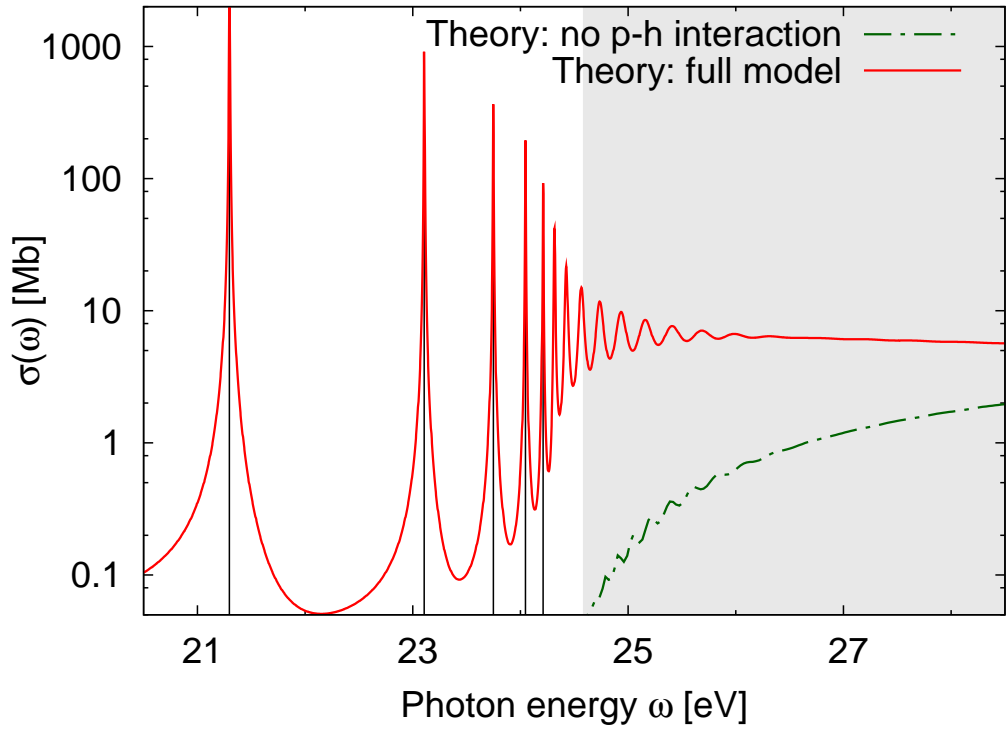}
  \caption{(Color online.) The calculated photoabsorption cross section $\sigma(\omega)$ of helium is plotted as a function of the incident photon energy $\omega$. Disabling the interaction of the excited electron and the corresponding hole (dot-dashed line), photoabsorption takes place only in the ionization continuum (shaded area). In the full model (solid line), Rydberg lines pertaining to the \mbox{$1s^2 \rightarrow 1s\,np\,(^1P_1)$} series emerge, the positions of the first five of which are marked by vertical lines.
  }
  \label{fig:herydb}
\end{center}
\end{figure*}

Illustrating this concept further, we present two photoabsorption spectra of helium calculated in the vicinity of the ionization threshold (see Fig.~\ref{fig:herydb}).
In the first calculation (dot-dashed line) we suppress the interaction of excited electron and hole. To this end we neglect \emph{all} $\bra{\Phi^a_{i}}\hat{V}^{ee}\ket{\Phi^b_{j}}$ matrix elements so that an excited electron ``experiences'' only the mean-field potential $\hat{V}^{MF}_n$, which is included in $\hat{H}_0$.
Since $\hat{V}^{MF}_n$ represents the \emph{neutral} atom it cannot account for the attractive Coulomb interaction between an excited electron and the hole. 
Without this attractive interaction no bound electron-hole pair can be formed.
This circumstance is reflected in the photoabsorption cross section, which features no bound-to-bound transition lines, but rises only as unbound electron-hole configurations become energetically accessible. (This is the case for the ionization continuum, marked by the shaded area in Fig.~\ref{fig:herydb}.)

In the second case (see Fig.~\ref{fig:herydb}, solid line) we include the interaction of electron and hole, which then gives rise to the \mbox{$1s^2 \rightarrow 1s\,np\,(^1P_1)$} Rydberg series in the photoabsorption spectrum.
Note that although Rydberg series consist---in principle---of infinitely many resonances, the number of lines in Fig.~\ref{fig:herydb} is actually finite due to the finite size of the numerical box.\cite{endnote5}

Using the positions of the lines $\omega_n$ as a benchmark, we can investigate the performance of TDCIS.
In Table~\ref{tab:quantumdefectHe} we list the first five transition energies and compare them to experimental data from Ref.~\onlinecite{NISTwebsite}.
The agreement between theory and experiment is found to be moderate for the first two lines, whereas it improves for transitions into higher excited states. This result instructively reflects the abilities and limitations of the TDCIS model. 
Describing an excited state of helium, the model accounts for an excited electron and the corresponding hole, which is generated in the 1s orbital. The hole---or respectively the remaining electron---is ``frozen'' in its state, because any relocation would correspond to a second excitation, which is not allowed in the CIS ansatz.
This rigid configuration within our model is insufficiently relaxed and unable to respond to the polarizing influence of the excited electron. This leads to a discrepancy between calculated and measured transition energies, which decreases towards higher excitations, as the electron becomes more distant and, therefore, its polarizing influence grows weaker.

From this example we can learn that higher-order excitations, going beyond the CIS approach, may be required to describe electron correlations, although the CIS wave-function ansatz (Eq.~\eqref{eq:TDCISwavefct}) captures the fundamental effects remarkably well.
In Appendix~\ref{app:forms} we outline a systematic approach to investigate the limitations of our model and thus gain an estimate of the significance of higher-order correlations.

\subsubsection*{Applications in other fields of physics}

Rydberg spectra, which are likewise affected by strong many-body correlations, can be found in molecular systems\cite{Sa-2002-BOOK_Rydberg_States_Spectroscopy,MaSc-2012-Rydberg_series_H2} and in solid-state physics, pertaining to excitonic phenomena.\cite{Li-1970-PhysEduc_Excitons} 
In the latter case, the picture of bound particle-hole excitations typically forms the basis of theoretical descriptions (see, for example, Ref.~\onlinecite{Ma-1996-BOOK_Introduction_solid_state_theory}).

For molecular systems, Rydberg spectroscopy has become a crucial tool to identify molecules and particularly to distinguish isomers.\cite{Sa-2002-BOOK_Rydberg_States_Spectroscopy,SHLi-2007-Rydberg_in_molecule}
Isomers are compounds that have the same molecular formula but different structure.
Chemically, isomers can have quite different behaviors and, therefore, it is important to differentiate them.

In contrast to atoms, molecular Rydberg states have not only an electronic character but also vibrational and rotational characters, where the latter two are highly sensitive to positions of the atoms within the molecules.
Each isomer has, therefore, unique Rydberg lines (position and strength) that make it possible to tell isomers apart.

\subsection{Fano resonances}
\label{sec:spec_fano}
As pointed out earlier, the ionization threshold of each atomic subshell is preceded by a series of resonances pertaining to discrete excited states of the atomic system.
The very first of these Rydberg series occurs below the first ionization threshold of the atom, while all subsequent series will be found above this ionization energy. The latter are thus embedded in the ionization continuum of more weakly bound subshells (cf.\ the 4s series of krypton between $20$~eV and $30$~eV in Fig.~\ref{fig:krover}).
Nevertheless, the electronic excitations associated with these latter series are bound states of the atomic system; as they do not exceed the specific ionization energy that is required for their respective subshell, their channel is said to be \emph{closed}.
These excitations can, however, still lead to an ionization of the atom, if the energy can be transferred from the closed channel into an open one by electron-electron interactions.
This effect is known as autoionization.

\begin{figure*}[ht!]
\begin{center}
  \rmpdfinfo
  \includegraphics[width=.8\textwidth]{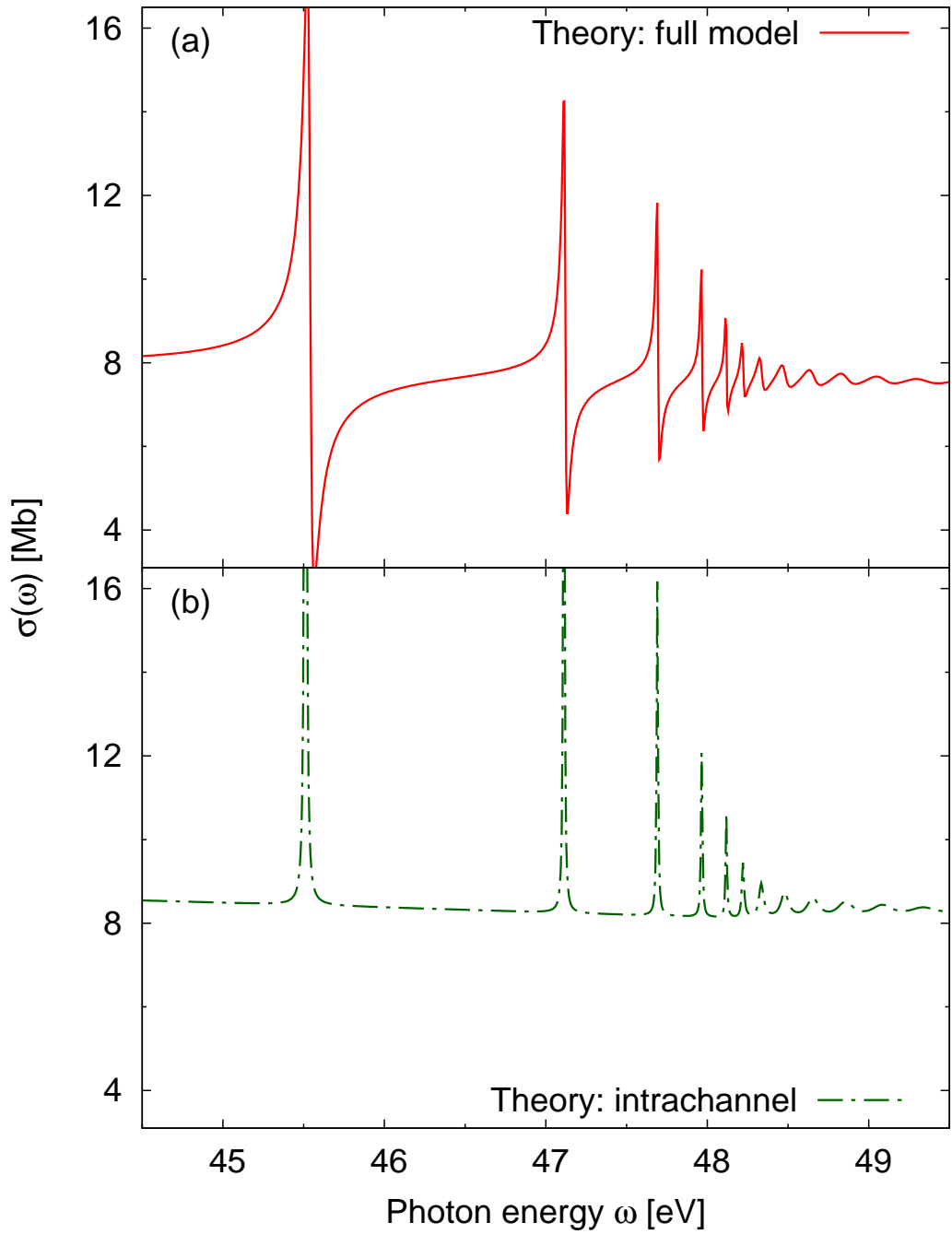}
  \caption{(Color online.) The calculated photoabsorption cross section $\sigma(\omega)$ of neon is plotted as a function of the incident photon energy $\omega$. Using the full model, the autoionizing Rydberg series \mbox{$2s^22p^6 \rightarrow 2s2p^6\,np\,(^1P_1)$} exhibits Fano resonances (panel (a)), whereas using the intrachannel model, the lines appear with a Lorentzian profile (panel (b)).
  }
  \label{fig:nefano}
\end{center}
\end{figure*}

An example of autoionizing resonances can be seen in panel (a) of Fig.~\ref{fig:nefano}, displaying the Rydberg series of the $2s$ subshell in neon.
The depicted resonance profiles appear noticeably different from the Lorentzian shape of common resonance phenomena, an observation that was first reported by Beutler\cite{Be-1935-ZSfPhysik_fanolinien} and subsequently explained by Fano.\cite{Fa-1935-NuovoCimento_fanolinien,Fa-1961-PhysRev.124.1866-fanolinien}
These features can be well reproduced (compare, for example, the results of Ref.~\onlinecite{CoMa-1967-PhysRev.155.26FanoInNeon}) and understood within the framework of TDCIS in terms of the interference of different ionization pathways.

In the vicinity of an autoionizing resonance, the energy of a photon ($\gamma$) suffices to ionize the atomic system (here neon) directly, by exciting an electron out of its outermost subshell into the continuum of unbound states, leaving a $2p$ hole: \mbox{$1s^22s^22p^6+\gamma \rightarrow 1s^22s^22p^5+e^-(\varepsilon)$},
where $e^-(\varepsilon)$ indicates a free electron of energy $\varepsilon$.
For photon energies close to an autoionizing resonance, the hole can likewise be created in the energetically deeper-lying $2s$ subshell, producing an excited---but bound---electron-hole pair, which appears as a Rydberg state of principal quantum number $n$: \mbox{$1s^22s^22p^6+\gamma \rightarrow 1s^22s^12p^6\,np$}.
Mediated by the Coulomb interaction between the electron and the hole, the latter can be promoted from the $2s$ into the $2p$ shell, transferring the excess energy to the excited electron.
In this way the interchannel coupling $\bra{\Phi^a_{i}}\hat{V}^{ee}\ket{\Phi^b_{j}}$ (cf.\  Section~\ref{sec:theo_TDCISwavefct}) mixes the closed and the open channel, leading to an autoionizing decay of the bound excitation $1s^22s^12p^6\,np$ into the ionized state $1s^22s^22p^5+e^-(\varepsilon)$.
These different ionization paths, which are indistinguishable by their final result ($1s^22s^22p^5+e^-(\varepsilon)$), interfere as their transition amplitudes add up. By tuning the photon energy, this interference is constructive on one side of the resonance and destructive on the other side, giving rise to the asymmetric profile of Fano resonances.

The significance of electron-electron interactions to this process can readily be demonstrated using TDCIS, as it allows one to systematically enable (full model) or disable (intrachannel model) electronic coupling of channels.
As can be seen from panel (b) in Fig.~\ref{fig:nefano}, the asymmetric profiles give way to symmetric resonances if the coupling of electronic channels is disabled. The bound excitations stemming from the $2s$ orbitals are prevented from autoionizing in this case and thus exhibit a ``normal'' Rydberg series.

\subsubsection*{Applications in other fields of physics}

Fano resonances are not restricted in their occurrence to atomic spectroscopy. They appear in a wide range of fields, whenever an intrinsically closed channel couples to a continuum.
The scattering properties of ultracold atoms can for example be influenced using externally tuned Fano resonances,\cite{InKe-nature1998-feshbachinBEC} allowing for the investigation of novel constellations of matter, like Efimov states.\cite{KrMa-2006-nature_EFIMOV}
In the context of scattering physics these Fano resonances are often referred to as Feshbach resonances.\cite{Fe-1962-unifiedNuclearResonances} 

In solid state physics, Fano resonances show up in a variety of processes.\cite{MiFl-RMP-2010} 
Besides photoelectron spectroscopy,\cite{ReHu-NJP-2005} which is closely related to photoionization spectroscopy, Fano resonances can also be seen in the conductivity of metallic systems~\cite{MaCh-Science-1998} and more generally in solid state systems that exhibit the Kondo effect.\cite{BuSt-PRL-2001,Ko-PTP-1964} 
The Kondo effect occurs in solids with lattice defects (impurities).
Electrons traveling through the solid can scatter from these impurity sites.
The interference between scattered and unscattered electrons affect the overall conductivity. 
Depending on the applied voltage, the currents of these different pathways interfere constructively or destructively, leading to an enhancement or reduction in the conductivity, respectively. 
The resulting behavior of the conductivity as a function of the applied voltage corresponds exactly to the Fano line shape.

Further examples for the occurrence of Fano resonances in complex systems can be found in Ref.~\onlinecite{SaGa-2008-creation_feshbach_resonance_molecule}, studying the creation of Feshbach resonances in the dissociation of a molecule, or Ref.~\onlinecite{MiFl-2008-REVIEW_Fano_in_solid_state}, which reviews Fano resonances in nanoscale structures.

\subsection{Cooper minimum}
\label{sec:spec_coop}
Up to now, we have studied the influence of many-electron effects on spectroscopic features involving transitions to excited bound states (Section~\ref{sec:spec_rydberg}) or autoionizing bound states (Section~\ref{sec:spec_fano}).
Also in the continuum, when the photon energy is large enough to directly ionize an atom, spectroscopic features appear that can be related to the electronic structure of the respective system. 

\begin{figure*}[ht!]
\begin{center}
  \rmpdfinfo
  \includegraphics[width=.8\textwidth]{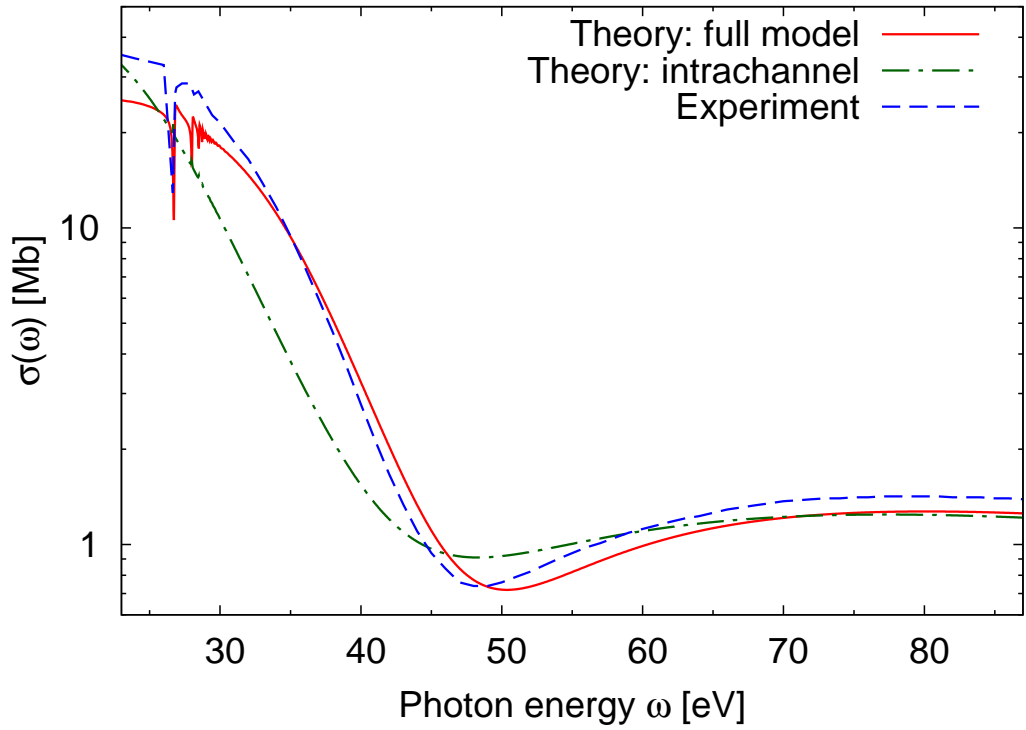}
  \caption{(Color online.) The photoabsorption cross section $\sigma(\omega)$ of argon, calculated with the full model (solid line) and the intrachannel model (dot-dashed line), is compared to the experimental results (dashed line) from Ref.~\onlinecite{SaSt-2002-JoElSp2002265}. A pronounced Cooper minimum is visible around $50$~eV.
  }
  \label{fig:arcoop}
\end{center}
\end{figure*}

In this section, we focus on an effect called the Cooper minimum.\cite{Co-1962-PhysRev.128.681Coopermin}
In Fig.~\ref{fig:arcoop} we show the pronounced Cooper minimum in the photoabsorption cross section of argon as an example.
Even though the existence of a Cooper minimum can be explained by an independent particle picture (i.e., a mean-field theory), the position as well as the shape of the Cooper minimum is sensitive to many-body effects (as one can see in Fig.~\ref{fig:arcoop}). 
Specifically, the TDCIS results with and without interchannel coupling are shown, together with experimental data.
By including the interchannel coupling effects, the position ($\approx 48$~eV) and depth ($\approx 0.7$~Mb) of the minimum is reproduced much better than in the case where only intrachannel coupling is allowed, which produces a too-shallow minimum. 
To improve the theoretical curve even further, higher order excitations (like double excitations) would have to be included, going beyond CIS.\cite{Starace-HandbuchPhysRadiation1982}

The origin of the Cooper minimum, which is not in itself a many-body effect, can be easily understood. 
The absorption of a photon changes the angular momentum by one unit. 
In the case of argon, for the photon energies shown in Fig.~\ref{fig:arcoop}, predominantly an electron from the $3p$~shell will be ionized.
The angular momentum, $l$, of the electron after absorbing a photon is, therefore, either $l=0$ or $l=2$, where the latter one is the dominant transition.\cite{FaCo-1968-RevModPhys.40.441} 
The Cooper minimum arises when the dominant transition matrix element undergoes a change of sign---i.e., it passes through zero---in the course of rising excitation energies. 

In most cases the total photoabsorption cross section will nevertheless be different from zero at the minimum position, due to contributions from the non-dominant excitation channels. 
According to Fano and Cooper,\cite{FaCo-1968-RevModPhys.40.441} minima can be excluded for orbitals that are radially nodeless ($1s$, $2p$, $3d$, $4f$, $\ldots$).
Photoionization from any other orbital may well exhibit a Cooper minimum. However, Fano and Cooper\cite{FaCo-1968-RevModPhys.40.441} also remark that the second rise of the cross section, and thus the distinctive shape of the minimum, can be largely obscured by other spectroscopic features.

\subsubsection*{Applications in other fields of physics}

Cooper minima have received recent attention, as evidence for their occurrence in high-harmonic-generation (HHG) spectra was found.\cite{WoNi-2008-PhysRevLett.102.103901_Cooper_min_HHG_Argon}
HHG is the most fundamental process in the emerging field of attosecond physics.
It is because of the HHG process that nowadays pulses shorter than 1~fs can be generated. 
The HHG process can be explained in terms of three steps:\cite{Co-PRL-1993,LeBa-PRL-1994} (1) an electron is ionized by a strong-field pulse; (2) the direction of the electric field of the pulse inverts and drives the electron back to the ionized system; and (3) the electron recombines with the ionized system and emits a high energy photon in the form of an attosecond pulse.

Particularly in the last step, the recombination, the electronic structure of the system plays an important role and strongly influences the HHG spectrum.
Since the mechanism of recombination is directly related to the mechanism of photoionization, the Cooper minimum appears also in the generation of attosecond pulses.
More generally, all electronic structure features (including many-body effects) influencing the photoionization cross section in the continuum do also influence the HHG process and in broader terms the world of attosecond physics.

Besides atomic systems, Cooper minima were likewise reported in the photoabsorption cross section of molecules by, for example, Carlson and coworkers,\cite{CaKr-1982-JChemPhys-Coopermin_mol} though the increased complexity of molecular systems inhibits the formulation of simple orbital rules.

\subsection{Giant resonance in xenon}
\label{sec:spec_gdre}
\begin{figure*}[ht!]
\begin{center}
  \rmpdfinfo
  \includegraphics[width=.8\textwidth]{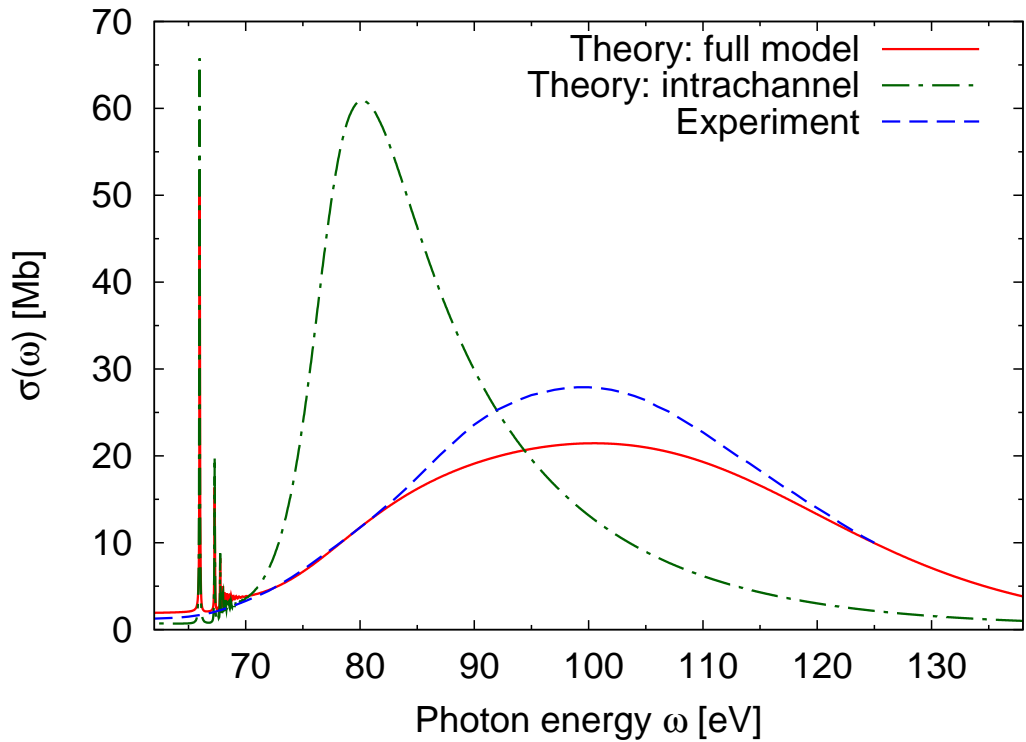}
  \caption{(Color online.) The photoabsorption cross section $\sigma(\omega)$ of xenon, calculated with the full model (solid line) and the intrachannel model (dot-dashed line), is compared to the experimental results (dashed line) from Ref.~\onlinecite{SaSt-2002-JoElSp2002265}. The intrachannel model is insufficient to reproduce the giant resonance around $100$~eV.}
  \label{fig:xegdre}
\end{center}
\end{figure*}

Another noteworthy spectroscopic feature is the giant resonance associated with the photoionization of the $4d$ subshell in xenon.
In 1964 Ederer\cite{Ed-1964-PhysRevLett.13.760_GDR} and independently Lukirskii and coworkers\cite{LuBr-1964-OptSpectrosk_GDRXenon_transl} discovered that the photoabsorption cross section of xenon exhibits an unusual shape above the $4d$ ionization threshold at $67.5$~eV.\cite{XDB}
Instead of the generic, monotonic decrease of the continuum cross section, an increase was observed, peaking around $\omega=100$~eV.  An extended resonance-like profile is formed between $\omega=70$~eV and $\omega=140$~eV, as shown in Fig.~\ref{fig:xegdre} (with the experimental data\cite{SaSt-2002-JoElSp2002265} plotted as a dashed line). This resonance differs decidedly from the previously discussed Rydberg and Fano resonances, because it does not pertain to the excitation of a bound excited state, but lies within the ionization continuum.

Like the Cooper minima discussed in the previous section, this giant resonance can be understood by using a single-electron picture, whereas the explanation of its precise shape, strength, and position requires many-electron correlations.\cite{Co-1964-PhysRevLett.13.762}
The latter may instructively be investigated using TDCIS.

Regarding a single electron from the $4d$ subshell of xenon, its photoexcitation will dominantly take place into $l=3$ excited states.
The accessibility of these states is, however, reduced for low excitation energies due to a radial potential barrier.\cite{Starace-HandbuchPhysRadiation1982}
With increasing photon energy the accessibility is improved, which leads to a growing ionization probability.
The highest ionization probability is reached when the energy of the excited electron is comparable to the height of the barrier.\cite{Starace-HandbuchPhysRadiation1982} 

We can retrace this picture of a single, independent electron using the TDCIS intrachannel model. This yields a result (see dot-dashed line in Fig.~\ref{fig:xegdre}) that disagrees with the experiment in shape, strength, and position, though it does predict a resonance-like appearance, resembling the single-electron curve presented by Cooper.\cite{Co-1964-PhysRevLett.13.762}
The overall profile of the resonance, however, is determined by strong collective effects within the valence shells of xenon, which evolve due to electron-electron interactions, as long as the excited electron is ``trapped'' behind the radial potential barrier. Amusia and Connerade suggest that these interactions form a plasmon-like coherent oscillation of at least all $10$ electrons in the $4d$ subshell.\cite{AmCo-2000-ProgRep_GDR}
Using the full model, wherein the interchannel couplings allow for a coherent superposition of all single excitations of the wave-function ansatz (Eq.~\eqref{eq:TDCISwavefct}),
we can capture parts of these collective dynamics, leading to decidedly better agreement with the experimental data (see the solid line in Fig.~\ref{fig:xegdre}). This underlines the importance of electronic correlations and the mixing of channels in photoionization processes.

\subsubsection*{Applications in other fields of physics}

Collective phenomena like the giant resonance are of importance in various physical disciplines. 
They were first observed in 1946 by Baldwin and Klaiber\cite{BaKl-1947-PhysRev.71.3_GDR_nuclei_first} for atomic nuclei, where they are likewise of relevance to astrophysical research.\cite{Kh-2007-GDR_astro} 
In atomic physics these resonances appear for a variety of elements\cite{Starace-HandbuchPhysRadiation1982} and also in molecular spectroscopy comparable structures were found, where they launched a long-lasting discussion on their potential to predict bond lengths.\cite{Pi-1999-ShapeResoMolecules} 

In solids and plasmas the collective motions of electrons, atoms, and both together are known as plasmons, phonons, and polarons, respectively.\cite{Kittel-2005-book}
The existence of plasmons can be seen in our daily life when we look at metals. 
Their shininess is a direct consequence of the collective motion of the electrons in the metal when irradiated by light.\cite{Fu-2011-book}
When the frequency of light is smaller than the characteristic frequency of the plasmons (known as the plasma frequency), the electrons oscillate collectively in phase with the light and prevent the light from entering the medium and reflect it from the surface. 
When the frequency is higher than the plasma frequency, the collective electron motion is too slow and cannot respond to the fast field oscillations.
As a result, the metal becomes transparent for these frequencies. 
For metals the plasma frequency is typically in the ultraviolet regime, so optical light is reflected, leading to the shiny appearance.

Furthermore, plasmons and the plasma frequency are highly sensitive to the electronic structure of the metal. 
This can be used to study effects like adsorption of material on metal surfaces. 
This is known as the surface plasmon resonance technique and is a research field of its own with many applications in biochemistry.\cite{Ho-2006-book,Pa-ABB-2006}

\section{\label{sec:conclusion} Conclusion}
%
%
Our aim in this article was to present atomic physics as a suitable starting point for introducing students to some of the basic penomena of many-body physics.
To this end we studied the impact of many-body effects upon the photoabsorption cross sections of noble gas atoms.
We demonstrated that such influences can be found in prominent spectroscopic features such as Rydberg series, Fano
resonances, Cooper minima, and giant resonances. 

In order to facilitate the understanding of the occurring many-body interactions we introduced the TDCIS model, which is conceptually simple, but allows for a systematic investigation of electronic correlations that go beyond a mean-field picture.
Regarding the good agreement of TDCIS results with experimental data, we can conclude that our intuitive model of particle-hole excitations captures a wide range of many-body effects in the studied closed-shell systems. The influence of higher-order excitations, which are not accounted for within our wave-function ansatz, is further elucidated in Appendix~\ref{app:forms}.

We want to emphasize that the conceptual
clearness of TDCIS in combination with the relative simplicity of atomic systems enables a comprehensible
introduction to basic many-body phenomena at early stages of the physics curriculum. This elementary introduction can be linked to several more advanced topics of many-body physics, as was outlined in the course of the article. 

Atomic physics holds further intriguing many-body phenomena, especially if attention is turned towards open-shell atoms.
For closed-shell systems we studied correlations only in the excited states. However, for open-shell atoms already the ground state belongs to a strongly correlated multiplet,\cite{Jo-2007-BOOK-Atomic_Structure_Theory} which must be represented by a superposition of multiple Slater determinants. 
These systems allow for a pedagogical introduction of ordering phenomena (cf.\ Hund's rules\cite{BrJo-2003-Atoms_Molecules_Hunds_rules}),
eventually linking to the study of magnetism\cite{Ma-2006-Magnetism_Simple,ZiPr-2002-Magnetism_Hubbard} or the popular Hubbard model.\cite{Ma-2000-Many_Particle_Hubbard,JoSt-2008-Mott_Insulator_Fermionic_nature}

\appendix

\section{Comparison of dipole forms}
\label{app:forms}

TDCIS has proved an adequate model to describe a wide range of phenomena, although its wave-function ansatz is limited to single electronic excitations. In the following, we outline how the necessarily insufficient description of higher order excitations can be quantified and used as an instructive indicator for the significance of higher order many-body effects.
We cannot gain such a measure from a direct comparison with experimental data, as this procedure would not single out deviations due to the neglected relativistic effects.
Therefore we shall rather employ two different representations of the photoabsorption cross section within the theoretical framework---the velocity and length forms---and judge their mutual disaccord.

\begin{figure*}[ht!]
\begin{center}
  \rmpdfinfo
  \includegraphics[width=.8\textwidth]{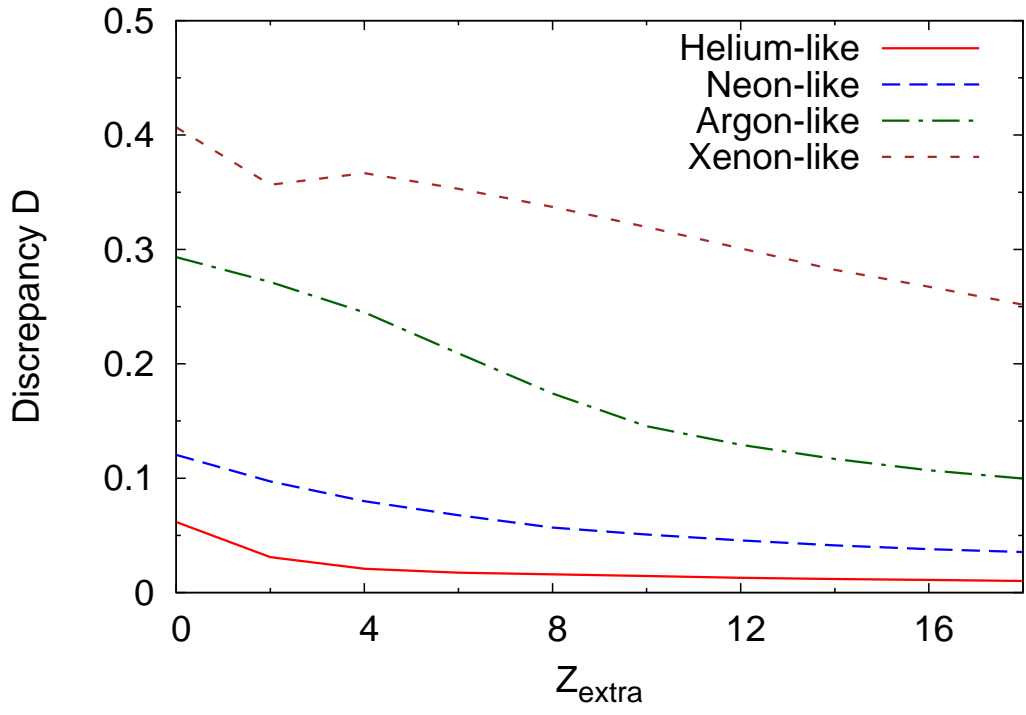}
  \caption{(Color online.) The discrepancy $D$ of velocity and length forms (Eq.~\eqref{eq:D}) is plotted as a function of additional nuclear charge $Z_{\text{extra}}$ for atomic systems with helium-like, neon-like, argon-like, and xenon-like electronic configurations. The neutral atoms have $Z_{\text{extra}}=0$.}
  \label{fig:MBSeries}
\end{center}
\end{figure*}

As early as 1945, Chandrasekhar\cite{Ch-1945-introduction_lv} pointed out that the photoabsorption cross section could be written in the velocity form, as we have done so far (cf. Eq.~\eqref{eq:tdpt_cross}),
\begin{align}
  \label{eq:tdpt_cross_v}
  \sigma_v(\omega)
  &=
  \frac{4 \pi^2}{\omega c} \sum_{F} \left| \bra{\Psi_F}\hat{P}\ket{\Phi_0} \right|^2 \delta(E_F - E_0 - \omega)
,
\end{align}
or the length form,
\begin{align}
  \label{eq:tdpt_cross_l}
  \sigma_l(\omega)
  &=
  \frac{4\pi^2\omega}{c} \sum_{F} \left| \bra{\Psi_F}\hat{Z}\ket{\Phi_0} \right|^2 \delta(E_F - E_0 - \omega )
,
\end{align}
with $\hat{Z}$ representing the $z$-components of the position operators in the abbreviated manner used before with the momentum operators (see Section~\ref{sec:theo_hamiltonian}). Equations~\eqref{eq:tdpt_cross_v}~and~\eqref{eq:tdpt_cross_l} are connected by the operator identity\cite{BeSa-2008-BOOK_Dover_Bethe_Salpeter}
%
%
\begin{align}
  \label{eq:comm_rela}
  \hat{P}
  &=
  - i [ \hat{Z},\hat{H} ].
\end{align}
Generally, Eq.~\eqref{eq:comm_rela} loses its necessary validity if the Hamiltonian employed is approximate or the basis set truncated (as in our case).
Then the equality of the velocity- and the length-form cross sections (Eqs.~\eqref{eq:tdpt_cross_v}~and~\eqref{eq:tdpt_cross_l}) is also broken---a circumstance that has caused a prolonged controversy over the ``right'' choice of form.\cite{St-1971-PhysRevA.3.1242_LV,CoMc-1972-ChemPhysLett_LV,Li-1978-PhysRevA.17.1939_LV,Ko-1979-PhysRevA.19.205_l_v_gauge_invariance_l}
Beyond the controversy, however, remains the fact that both forms will give the same result if the description of the system is exact.
Therefore we can estimate the accuracy of our approximation and hence the significance of higher order many-body contributions by using the normed discrepancy between these two forms:
\begin{align}
  \label{eq:D}
  D
  &=
  \frac{\int d\omega \left| \sigma_v(\omega) - \sigma_l(\omega) \right|}{\int d\omega \left( \sigma_v(\omega) + \sigma_l(\omega) \right)}.
\end{align}
In Fig.~\ref{fig:MBSeries} we show $D$ evaluated\cite{endnote6} for atomic systems with helium-like, neon-like, argon-like, and xenon-like electronic configurations.  In addition to the results for the neutral atoms, we also show $D$ for atomic nuclei with additional charge ($Z_{\text{extra}}$); a similar approach can be found in Ref.~\onlinecite{QuLa-1984-AustJPhys_LVIso}.

We observe that with an increasing number of electrons incorporated in an atomic system, the discrepancy of the cross section forms grows.
This is readily understandable, as in larger atomic species, the outer electrons are weakly bound, yet strongly interacting, which gives rise to higher-order collective phenomena.
For systems with an increasing additional nuclear charge, this discrepancy between velocity and length forms decreases again, indicating a decline of the many-body contributions.
This second observation is plausible, as with higher nuclear charge the single-electron central potential (cf.\ Eq.~\eqref{eq:Hamiltonian2}) increases in importance relative to the electronic many-body interactions. Thus the described system becomes more hydrogen-like with growing $Z_{\text{extra}}$, incorporating strongly bound and comparatively weakly interacting electrons. Hence, the TDCIS description in terms of single excitations becomes progressively accurate for high $Z_{\text{extra}}$, which is reflected in the observable convergence trend of velocity- and length-form cross sections.

%
%
\section{Alternative calculation of the photoabsorption cross section}
\label{app:cross}
The approach to study photoabsorption based on the depopulation of the ground state (see Section~\ref{sec:theo_TDSE}) is instructive and universal, but largely inappropriate for the calculation of a weak-field photoabsorption cross section $\sigma(\omega)$. To obtain satisfactory energetic resolution with this method, a spectrum has to be sampled by a multitude of pulses, each requiring an individual solution to Eqs.~\eqref{eq:eom}.
A notably more efficient scheme to find $\sigma(\omega)$ can be arrived at if the familiar (see, e.g., Ref.~\onlinecite{Sa-2009-J.Phys.B_phd_tutorial_XRAY}) expression for the photoabsorption cross section derived from first-order time-dependent perturbation theory is employed:
\begin{align}
  \label{eq:tdpt_cross}
  \sigma(\omega)
  &=
  \frac{4 \pi^2}{\omega c} \sum_{F} \left| \bra{\Psi_F}\hat{P}\ket{\Phi_0} \right|^2 \delta(E_F - E_0 - \omega).
\end{align}
In Eq.~\eqref{eq:tdpt_cross}, $\ket{\Psi_F}$ are eigenstates of the full, yet field-free Hamiltonian, with $E_F$ their corresponding energy eigenvalues; $E_0$ denotes the ground-state energy and $c$ the speed of light in vacuo. Following Tong and coworkers,\cite{ToTo-2010-PhysRevA.81.063403} we restate Eq.~\eqref{eq:tdpt_cross} as essentially the inverse Fourier transform of a correlation function $C(t)$, yielding
\begin{align}
  \label{eq:corrfct_cross}
  \sigma(\omega)
  &=
  \frac{4 \pi }{\omega c} \text{Re}\left[{\int_{0}^{\infty}\!\!\!dt ~C(t)~e^{i \omega t}}\right]
  .
\end{align}
Here, the correlation function has to be taken as the overlap between an initially dipole-disturbed ground state $\ket{{\Psi}^{\prime}(0)}=\hat P\ket{\Phi_0}$ and its field-free propagated self $\ket{{\Psi}^{\prime}(t)}$, hence $C(t)=\braket{{\Psi}^{\prime}(0)}{{\Psi}^{\prime}(t)}$.
A detailed derivation of Eq.~\eqref{eq:corrfct_cross} can be found in Ref.~\onlinecite{SchatzRatner_book}.
With this the numerical effort reduces to solving only once the field-free ($A(t)\equiv 0$) set of Eqs.~\eqref{eq:eom}:
\begin{subequations}
\label{eq:eom_ff}
\begin{eqnarray}
\label{eq:eom_alpha0_ff}
  i\dot\alpha_0(t)
  &=&
  0,
\\
\label{eq:eom_alpha_ai_ff}
  i\dot\alpha^a_{i}(t)
  &=&
  (\varepsilon_a\!-\!\varepsilon_i) \, \alpha^a_{i}(t)
  +
  \sum_{b,j}
 \bra{\Phi^a_{i}}\hat{V}^{ee}\ket{\Phi^b_{j}}
  \alpha^b_{j}(t),
\end{eqnarray}
\end{subequations}
with the appropriate initial conditions (i.e., $\alpha_0(0)=0$ and $\alpha^a_{i}(0)=\bra{\Phi^a_{i}}\hat P\ket{\Phi_0}$). This procedure reconstructs the entire cross section $\sigma(\omega)$ from one single propagation.
The correlation function can to this end be expressed through $\alpha^a_{i}(t)$:
\begin{align}
  \label{eq:corrfct_expansion}
  C(t)
  &=
  \sum_{a,i} \alpha^a_{i}(t)  \bra{\Phi_0}\hat P\ket{\Phi^a_{i}}
  .
\end{align}

\begin{acknowledgments}
This work has been supported by the Deutsche Forschungsgemeinschaft (DFG) 
under grant No.\ SFB 925/A5.
\end{acknowledgments}


\begin{thebibliography}{76}
\bibitem{endnote1} Employing atomic units, the cross section is given in terms of the Bohr radius squared $a_0^2$, which corresponds to $2.8 \,\, 10^{-21}$~m$^2$. In a more commonly used unit, this is $28$~Mb.
\bibitem{endnote2} The number of photons per unit area can be calculated as the time-integrated photon flux $N_\gamma=c\int dt \dot{A}(t)^2 / 4\pi\bar{\omega}$, where $\bar{\omega}$ is the mean photon energy of the pulse and $c$ the speed of light in vacuo.
\bibitem{endnote4} In order to improve the comparability of TDCIS results and experimental data, we adjusted the HF orbital energies to match the experimental binding energies from Ref.~\onlinecite{XDB}.
\bibitem{endnote5} The atom is simulated inside a finite numerical box, the boundaries of which conflict with the spatial extent of realistic highly excited states. By compressing these states into the box, which thereby acts as a confining potential-well, the states are shifted upwards in energy, across the ionization limit.
\bibitem{endnote6} Ideally the integration of the cross sections should be carried out from $0$ to $\infty$. However, our calculations were limited to the outer subshells and therefore the upper integration boundary was fixed relative to the threshold of the most strongly bound active orbital.
\end{thebibliography}
%
%


\clearpage

\begin{table*}[ht!]
  \caption{\label{tab:quantumdefectHe} Energies $\omega_n$ of the first five transition lines ($n=2,\ldots,6$) in helium as calculated with TDCIS are given and compared to experimental data from Ref.~\onlinecite{NISTwebsite}.}
  \begin{ruledtabular}
  \begin{tabular}{l c c}
      & Theory & Experiment \\
    $n$   & Line pos. $\omega_n$[eV] &Line pos. $\omega_n$[eV] \\
  \hline
    2 & 21.2967  & 21.2180\\
    3 & 23.1057 & 23.0870\\
    4 & 23.7480 & 23.7421\\
    5 & 24.0471 & 24.0458\\
    6 & 24.2111 & 24.2110\\
  \end{tabular}
  \end{ruledtabular}
\end{table*}

\end{document}